\documentclass [12pt]{article}

\begin{document}

\begin{center}
\large\textbf{Area spectra of near extremal black holes}
\end{center}

\begin{center}
 Deyou Chen\textsuperscript{a,}\textsuperscript{b,}\footnote
{E-mail: dchen@cfa.harvard.edu},\quad Haitang Yang
\textsuperscript{a,}\footnote {E-mail: hyanga@uestc.edu.cn} and
Xiaotao Zu\textsuperscript{a}
\end{center}
 \begin{center}
\textsuperscript{a}{School of Physical Electronics, University of
Electronic Science and
 Technology of China, Chengdu,  Sichuan 610054, China}\\
\textsuperscript{b}{Harvard-Smithsonian Center for Astrophysics,
60
Garden Street, Cambridge, MA 02138, USA}\\
\end{center}

\textbf{Abstract:} Motivated by Maggiore's new interpretation of
quasinormal modes, starting from the first law of thermodynamics of
black holes, we investigate area spectra of a near extremal
Schwarzschild de sitter black hole and a higher dimensional near
extremal Reissner-Nordstrom de sitter black hole. We show that the
area spectra of all these black holes are equally spaced and
irrelevant to the parameters of black holes.

\textbf{1. Introduction}

It is widely believed that black holes play a significant role in
understanding the theory of quantum gravity. In 1970s, Bekenstein
discussed the quantum effects of black holes \cite{JDB}. Supposing
that the horizon area of a nonextreme black hole behaves as a
classical adiabatic invariant which corresponds to a quantum
entity with discrete spectrum, Bekenstein proposed the horizon
area should be quantized. The quantized area spectrum has the
following form

\begin{equation}
A_n = \gamma n\hbar ,(n = 1,2,3 \cdots ),
\end{equation}

\noindent where $\gamma $ is a dimensionless constant and $\hbar $
is related to Planck length as $l_p^2 = \frac{\hbar G}{c^3} $. In
this paper, we set $c = G = 1$. There are some works to quantize
the area of black holes and the main difference is the value of
$\gamma $. It is known that when a classical black hole is
perturbed by an exterior field, the solution of the perturbational
wave equation is defined as quasinormal modes (QNMs) and its
relaxation is governed by a set of the modes with complex
frequencies. This behavior has the same characteristic as that of
a damped harmonic oscillator \cite{MM}. QNMs of black holes plays
an important role not only in gravitational wave astrophysics
\cite{KS} but also in the context of the AdS/CFT conjecture
\cite{EW,RAK}. There is a great deal of researching on QNMs. Hod
found that the area spectrum can be obtained from the QNMs of the
black hole \cite {SH}. In his work, by applying Bohr's
correspondence principle on the ringing frequencies which
characterize a black hole, the missing ``link'' was found and the
area spacing (that is $4l_p^2 \ln j$, here  $j  = 3 )$ of a
Schwarzschild black hole was derived when the real part of the
QNMs introduced. This value of area spacing is consistent with
both the area-entropy thermodynamic relation and  statistical
physics arguments. Later Dreyer derived the area spectrum from the
quasinormal modes frequency from the perspective of loop quantum
gravity \cite{OD}. The value of $\gamma $ is in coincidence with
that derived by Hod. Based on work of Hod and Dreyer, Kunstatter
succeeded in deriving the Bekenstein-Hawking entropy spectrum for
higher dimensional spherically symmetrical black holes, with
semi-classical arguments\cite{GK}. In his work, the quantity

\begin{equation}
I = \int {\frac{dE}{\omega \left( E \right)}}
\end{equation}

\noindent is treated as an adiabatic invariant, where $E$ and
$\omega \left( E \right)$ denote the energy of a system and
vibrational frequency, respectively. Applying Bohr-Sommerfeld
quantization, one can get an equally spaced spectrum in the
semi-classical (large $n)$ limit, namely $ I = n\hbar $ .
Subsequently more progress are achieved.

Taking Hod's conjecture into account, Maggiore very recently put
forward that in the semiclassical limit, the area spectrum of a
black hole ought to be determined by the asymptotic value of a
physical frequency of the QNMs, defined as \cite{MM}:

\begin{equation}
\omega _n = \sqrt {\omega _R^2 + \omega _I^2 } ,
\end{equation}

\noindent This offers the area spectrum of black holes a new
explanation. Applied this interpretation to the Schwarzschild
spacetime, he concludes that the area spectrum of the horizon is
equally spaced and is quantized in units $\Delta A = \gamma l_p^2 $
with $\gamma = 8\pi $, different from that obtained by Hod, Dreyer
and Kunstatter. Based on this new interpretation, there are much
work appeared \cite{ECV,AJMM,KPS,DG,WLLR,SF,KR,BMV,ALO,YSM,
Wei:2010yx,KSN,GPS}. For example, in the research on the area
spectrum of slowly rotating Kerr black holes, Vagenas found the area
spectrum is not equally spaced \cite{ECV}. Lopez-Ortega has studied
the area spectrum of the higher dimensional de Sitter spacetime and
spherically symmetrical black hole in small charge limit \cite{ALO}.
Area spectra of (2+1)-dimensional black holes and the large AdS
black holes have been investigated by Fernando and Wei
\cite{WLLR,SF}. All of these results are based on the viewpoint that
the horizon area of nonextremal black holes behaves as a classical
adiabatic invariant. There are also some achievements on the area
spectra of extreme black holes \cite{SV}.

The purpose of this paper is to combine Maggiore's new
interpretation and the first law of thermodynamics to investigate
the area spectra of near extremal black holes in de sitter space.
We discuss a near extremal Schwarzschild de Sitter black hole and
a higher dimensional near extremal Reissner-Nordstrom de sitter
black hole. The results show that the area spectra are equally
spaced and irrelevant to the parameters of the black holes.

The paper is organized as follows. We investigate the area spectrum
of a Schwarzschild de Sitter black hole near extreme case by
combining the new interpretation of Maggiore and thermodynamical
properties in sect. 2.  The area spectrum of a higher dimensional
Reissner-Nordstrom black hole near extreme case is investigated in
Sect.3. Sect. 4 contains some discussion and conclusion.

\textbf{2. Area spectrum of a near extremal Schwarzschild de Sitter
black hole}

In this section, following Maggiore's work, we discuss the area
spectrum of a Schwarzschild de Sitter black hole near extreme case
from thermodynamical properties of black holes. The metric of the
Schwarzschild-de Sitter black hole is given by

\begin{equation}
ds^2 = - f\left( r \right)dt^2 + f^{ - 1}\left( r \right)dr^2 + r^2\left(
{d\theta ^2 + \sin ^2\theta d\phi ^2} \right),
\end{equation}

\noindent with
\[
f\left( r \right) = 1 - \frac{2M}{r} - \frac{r^2}{a^2},
\]

\noindent where $M$ represents the physical mass of the black hole
and the parameter $a$ is related to the cosmological constant $
\Lambda $ by $a^2 = \frac{3}{\Lambda }$. For $f\left( r \right) =
0$, there are three roots ( $r_c, r_h, r_0$), corresponding to the
cosmological horizon ($r = r_c )$, the event horizon ($r = r_h <
r_c )$ and a negative one without physical interpretation ($r =
r_0 )$, respectively. The Hawking temperature is  $T =
\frac{\kappa_h }{2\pi }$, where $\kappa_h$ represents the surface
gravity at the event horizon

\begin{equation}
\kappa _h = \frac{\left( {r_c - r_h } \right)\left( {r_h - r_0 }
\right)}{2a^2r_h }.
\end{equation}

\noindent Near the extreme case, the cosmological horizon is very
close to the event horizon and the surface gravity of the event
horizon can be approximated as \cite{CL}

\begin{equation}
\kappa _h \approx \frac{r_c - r_h }{2a^2r_h }.
\end{equation}

\noindent As explained in the first section, once the QNMs are
given, one can employ the first law of thermodynamics to calculate
the area spectrum. The QNMs of the near extremal Schwarzschild de
Sitter black hole was derived in \cite{CL},

\begin{equation}
\omega = \kappa _h \left[ {\sqrt {V_0 - \frac{1}{4}} - i\left( {n +
\frac{1}{2}} \right)} \right],
\end{equation}

\noindent with $V_0 = l\left( {l + 1} \right)$ for scalar as well
as $U(1)$ vector perturbations and $V_0 = \left( {l + 2}
\right)\left( {l - 1} \right)$ for gravitational perturbations.
$l$ stands for the angular quantum number and $n = 0,1,2,3 \cdots
$. The gravitational perturbations with $l = 2,3$ agree with the
results of Moss and Norman \cite{MN}. In Hod's work, the real part
of the mode was adopted to derive area spectra of black holes. In
this paper, following Maggiore's arguments, we are concerned with
the imaginary part in the large $n$ limit. Maggiore argued that
the physical frequency is related to the real part and imaginary
part. As $n \to \infty $, the imaginary part plays a dominant role
in eq. (7), then the physical frequency can be expressed as

\begin{equation}
\omega _n = \left| {\omega _I } \right| = n\kappa _h .
\end{equation}

When a black hole is regarded as a damped harmonic oscillator, the
change of energy  is related to the change of the physical
frequency of the harmonic oscillator, namely $\Delta E = \hbar
\Delta \omega _n $. For the Schwarzschild de Sitter black hole,
the energy is its ADM mass. Therefore we can get

\begin{equation}
\Delta M = \Delta E = \hbar \Delta \omega _n .
\end{equation}

In literature, the area spectra of nonextreme black holes were
studied by assuming that the horizon area behaves as an classical
adiabatic invariant. The action is written as $I = \int
{\frac{dE}{\omega \left( E \right)}} $ for a system with energy
$E$ and vibrational frequency $ \omega \left( E \right)$ in
Kunstatter's work \cite{GK}.  For the black hole near extreme case
we are considering here, we introduce the first law of
thermodynamics of the black hole to relate the variances of energy
and entropy. The first law is

\begin{equation}
dM = T \cdot dS,
\end{equation}

\noindent with finite  form

\begin{equation}
\Delta S = \textstyle{{\Delta M} \over T}.
\end{equation}

\noindent Combining with Eqs. (8), (9) and (11), we can get the
change of the entropy spectrum,

\begin{equation}
\Delta S = \textstyle{{\hbar \Delta \omega _n } \over T}
\end{equation}

\[
= 2\pi \hbar ,
\]

 \noindent where the relation $\Delta \omega _n =
\omega _n - \omega _{n - 1} = n \kappa _h - (n-1) \kappa _h =
\kappa _h $ and the Hawking temperature at the event horizon $ T =
\frac{\kappa _ {h} }{2\pi }$ were used in the last equality. It is
easy to see that the entropy spectrum has the following form $S =
2\pi \hbar n + C$, where $C$ is a constant, irrelevant to our
problem. According the Bekenstein-Hawking area-entropy relation,
one can easily obtain $ A = 8\pi \hbar n + 4C$. Thus

\begin{equation}
\Delta A = 8\pi \hbar ,
\end{equation}

\noindent which shows the area spectrum of the near extremal
Schwarzschild de sitter black hole is equally spaced. This is in
consistence with recent results and shows Maggiore's idea can be
extended to the near extremal de sitter black holes.

\textbf{3. Area spectrum of a higher dimensional near extremal
Reissner-Nordstrom de Sitter black hole}

In this section, we calculate the area spectrum of the charged de
sitter black hole for near extreme case.  The metric of the
Reissner-Nordstrom de sitter black hole is given by the line
element

\begin{equation}
ds^2 = - f\left( r \right)dt^2 + f^{ - 1}\left( r \right)dr^2 +
r^2d\Omega _n^2 ,
\end{equation}

\noindent with

\[
f\left( r \right) = 1 - \lambda r^2 - \frac{2M}{r^{n - 1}} +
\frac{q^2}{r^{2n - 2}},
\]

\noindent where $d\Omega _n^2 $ is the line element of a $n$
sphere. The parameters $M$ and $q$ are  the physical mass  and
electric charge of the black hole respectively \cite{KI}. $\lambda
$ is related to the cosmological constant. There are three
positive roots for $f\left( r \right) = 0$, representing  the
Cauchy horizon $r_a $, the event horizon $r_h $ and the
cosmological horizon $r_c $, respectively. Near extreme case, the
cosmological horizon is very close to the event horizon. Then the
surface gravity associated the event horizon is

\begin{equation}
\kappa _h \approx \frac{\left( {r_c - r_h } \right)\left( {n - 1}
\right)}{2r_h^2 }\left( {1 - nq^2} \right).
\end{equation}

\noindent The QNMs obtained by Cardosc \cite{CLM} is

\begin{equation}
\omega = \kappa _h \left[ {\sqrt {\frac{V_0 }{\kappa _h^2 } -
\frac{1}{4}} - i\left( {n + \frac{1}{2}} \right)} \right],
\end{equation}

\noindent where $V_0 = V_{S\pm } \left( {r_h } \right)\cosh \left(
{\kappa _h r_\ast } \right)^2$ and $n = 0,1,2,3 \cdots $. Parallel
to the strategy in section two, we know the area spectrum is
related to the physical frequency. From the first law of
thermodynamics of the black hole, we can get

\begin{equation}
dM - \Phi \cdot dQ - \Omega \cdot dJ = T \cdot dS.
\end{equation}

\noindent Since the black hole does not rotate, one has $\Omega
\cdot dJ =0$. The change of energy of the black hole is given by
$dE=dM - \Phi \cdot dQ $. Therefore, at the event horizon, one
obtains

\begin{equation}
T \cdot \Delta S = \Delta M - \Phi \cdot \Delta Q = \Delta E = \hbar
\Delta \omega _n ,
\end{equation}

\noindent where the physical frequency $\omega _n = \sqrt {\omega
_R^2 + \omega _I^2 } $ is obtained from eq.(16). When $n \gg 1$, the
frequency is mainly relied on the imaginary part of the QNMs, namely
$\omega _n = \left| {\omega _I } \right|_n$, and there is $\Delta
\omega _n = \left| {\omega _I } \right|_n - \left| {\omega _I }
\right|_{n - 1} = \kappa _h $. From eq. (18), the change of the
entropy spectrum can be obtained as

\begin{equation}
\Delta S = \textstyle{{\hbar \Delta \omega _n } \over T} = 2\pi
\hbar .
\end{equation}
By the same procedure as in section two, it is straightforward to
find

\begin{equation}
\Delta A = 8\pi \hbar ,
\end{equation}

\noindent which shows the area spectrum of the higher dimensional
Reissner-Nordstrom de sitter black hole near extreme case is equally
spaced. This is in accordance with the results of charged and higher
black holes \cite{WLLR,ALO}.

\textbf{4. Discussion and Conclusion}

In this paper, from thermodynamical properties of the black holes,
by adopting Maggiore's new interpretation of QNMs, we investigated
entropy spectra and area spectra of the near extremal
Schwarzschild de sitter black hole and the higher dimensional near
extremal Reissner-Nordstrom de sitter black hole. The results show
that the area spectra are all equally spaced. In
literature\cite{ECV}, there is a logarithmic term appeared in the
area spectrum of the slowly rotating Kerr black hole, where
Vagenas explained it as the logarithmic correction. When the
logarithmic term becomes dominant in some certain condition, it
introduces difficulty to the interpretation of entropy spectrum.
In fact, the logarithmic term can be avoided appearing in the area
spectrum, as addressed in the work of Medved and Myung
respectively \cite{AJMM,YSM}. When applied to the slowly rotating
Kerr black hole, our argument is parallel to that suggested by
Medved \cite{AJMM}. Therefore, the area spectrum of the slowly
rotating Kerr black hole is also equally spaced in the limit $J/M
\to 0$ where $J$ stands for the angular momentum.

In conclusion, we investigated the area spectrum of near extremal
black holes and found that the area spectra of these black holes are
all equally spaced and irrelevant to the parameters of these black
holes.

\textbf{Note added}: When we are in the final stage of writing the
manuscript, paper \cite{Li:2010vh} appears in the preprint
archive. The authors discuss the area spectrum of a near extreme
Schwarzschild de sitter black hole by introducing the adiabatic
invariant, which is different from our method.

\section*{Acknowledgments}
This work is supported in part by NSFC (Grant No.10705008) and
NCET.

\end{document}